\documentclass[preprint,floatfix] {revtex4} 
\newcommand{\rvec}{\mathrm {\mathbf {r}}} 
\usepackage{graphicx}
\usepackage{subfigure}
\begin{document}

\title{A new density functional method for electronic structure calculation of atoms and molecules}
\author{Amlan K.\ Roy}
\email{akroy@iiserkol.ac.in, akroy@chem.ucla.edu} 
\affiliation{Division of Chemical Sciences, Indian Institute of Science Education and \\
Research (IISER), Block HC, Sector III, Salt Lake, Kolkata-700106, India.}

\begin{abstract}
In recent years, density functional theory (DFT) has emerged as one of the most successful and 
powerful approaches in electronic structure calculation of atoms, molecules and solids, as evidenced 
from burgeoning research activities in this direction. 
This chapter concerns with the recent development of a new DFT methodology 
for accurate, reliable prediction of many-electron systems. Background, need for such a scheme, 
major difficulties encountered, as well as their potential remedies are discussed at some length. 
Within the 
realm of non relativistic Hohenberg-Kohn-Sham (HKS) DFT and making use of the familiar LCAO-MO principle, 
relevant KS eigenvalue problem is solved numerically. Unlike the commonly used atom-centered grid (ACG), 
here we employ a 3D cartesian 
coordinate grid (CCG) to build atom-centered localized basis set, electron density, as well as 
all the two-body potentials directly on grid. The Hartree potential is computed through 
a Fourier convolution technique via a decomposition in 
terms of short- and long-range interactions. Feasibility and viability of our proposed scheme is 
demonstrated for a series of chemical systems; first with homogeneous, local-density-approximated 
XC functionals followed by non-local, gradient- and Laplacian-dependent functionals. 
A detailed, systematic analysis on obtained results relevant to quantum chemistry, are made, 
\emph{for the first time}, using CCG, which clearly illustrates the significance of this alternative 
method in the present context. Quantities such as component energies, total energies, ionization 
energies, potential energy curve, atomization energies, etc., are addressed for pseudopotential 
calculations, along with a thorough comparison with literature data, wherever possible. Finally, 
some words on the future and prospect of this method are mentioned. In summary, we have presented a 
new CCG-based \emph{variational} DFT method for accurate, dependable calculation of atoms and 
molecules. 
\end{abstract}
\maketitle

\section{Introduction}
With continuous rapid advances in sophisticated methodologies, techniques, algorithms as well as
ever-increasing growth of powerful computers, attempts to study electronic \emph{structure,
properties} of atoms, molecules, solids, clusters by direct \emph{ab initio} solution of 
many-body Schr{\"o}dinger equation has received an enormous amount of stimulus in the past several 
decades. The great challenge of solving this equation for realistic systems has opened up whole 
new avenues 
in chemistry, as, in almost all cases, no \emph{exact} analytic solution could be obtained (except in a small 
number of simple cases); hence recourse must be taken to approximate methods. Obviously, the ultimate goal 
is to achieve an accuracy obtained in the best experimental result using optimal computational resources. 
Now it is no more an unrealistic contemplation to think of highly accurate quantum mechanical 
calculations of molecules having 100 electrons or more, and quantum chemistry has been a full-fledged,
firmly established discipline for some time now.

Apart from the ground-state electronic structure (in terms of wave functions and densities), a 
theoretician ought to be able to tackle a multitude of other important aspects, such as excited 
states, spectroscopic term values, oscillator strengths for transitions, fine and hyperfine 
structure of energy levels due to relativistic corrections to Hamiltonian and so on. 
Molecular case, however, poses more overwhelming and spectacular challenges; for in addition to 
the physical properties mentioned (applicable to atoms as well), one has also needs to address a host 
of certain unique \emph{chemical} properties specific to them. For example, a molecule has a well-defined 
``geometry", which 
requires specification of various bond lengths, angles etc. Some other quantities of interest 
include, for example, force constants for stretching, bending (molecular vibration), energy needed 
to break a particular bond or to dissociate the molecules into fragments, nature of interaction 
between molecules, chemical reactions which may occur when they come into close proximity 
(synthetic and decomposition pathways), absolute and relative interaction energies, dipole and 
higher multipole moments, cross section for collision with other particles, stability of such a 
system (through second derivatives of energy with respect to the spatial coordinates, the Hessian 
matrix), the properties revealed by various modern spectroscopic methods (including electron spin 
resonance, nuclear magnetic resonance, nuclear quadrupole resonance, photoelectron spectroscopy, 
ro-vibrational spectroscopy, electronic spectroscopy) etc. 

Demands on the theory is no less severe. Considering the fact that energy required to break a bond 
is roughly of the order of 5 eV, while total electronic energy may be thousands of times higher, 
estimation of dissociation energy as a difference of two such large quantities, is clearly a 
non-trivial exercise. Furthermore, the elegant methods of angular momentum applicable to atomic systems 
can no longer be used for general polyatomic molecules which, in general, lack symmetry. Besides, we are 
interested in both static and dynamic situations. In all cases, the computer time and resources 
dramatically increase with system size in consideration. Thus a whole echelon of formalisms, ranging 
from highly accurate to very approximate, strive to achieve the best trade-off between accuracy and 
computational cost. The former \emph{ab initio} method is feasible only for relatively smaller 
systems, as these are based entirely on theory from first principles. The other, often less accurate 
and less reliable, empirical or semi-empirical methods usually employ experimental results to approximate 
some elements of the underlying theory. All these fall into the general gamut of modern computational
chemistry. A large number of excellent reviews, textbooks, monographs are available on the subject. 
Some of the recent ones are given here \cite{yarkony95,szabo96,simon97,kohn99,helgaker00,young01, 
lewars03,cramer04,martin04,hoffman07,jensen07,kaisas07}.

Two distinct lines of attack may be visualized. In the ``conventional" approaches (rooted in 
early works of Slater~\cite{slater60} on atoms, molecules), direct approximate solution of 
Schr\"{o}dinger equation is sought through carefully designed variational wave functions with the aid 
of most powerful computers available. In the indirect, second line of development, calculation of 
wave function is avoided by deriving expressions for observable themselves, namely electron 
density, giving birth to the versatile and popular density functional theory (DFT) \cite{march87,parr89, 
jones89,chong95,seminario96,joulbert98,dobson98,koch01,parr02,fiolhais03,gidopoulos03}. Last few 
decades have witnessed a proliferation of DFT-based methodologies for electronic calculations of a 
broad range of systems including atoms, molecules, condensed phases, materials science, etc. 
With diverse advancements in computer technology, this allows larger systems of physical and chemical 
interest to be approached than are ordinarily accessible within the domains of traditional methods, 
and this continues to grow at a steady pace. An 
overwhelming variety of important challenging problems are addressed ranging from to study properties 
of doped fullerene superconductor or a solid semiconductor; structure, modeling and docking of a 
peptide; catalytic properties of a zeolite or a surface layer; predict, design reaction pathways 
leading to the desired pharmaceutical product; electronic as well as spin-dependent (spintronic) 
transport properties of molecules; modeling of intermolecular forces and potential energy surfaces 
at the molecular level, etc. An extensive amount of successful applications have been made to a 
varied range of molecular studies as well; such as prediction of physicochemical properties like equilibrium 
geometries, harmonic vibrational frequencies, polarizabilities, hyperpolarizabilities, dissociation 
energies, stability, transition states, weak interactions like H-bonds, reaction pathways, etc. 
Indeed it is no more an overstatement that today's electronic structure calculations of materials is 
principally dominated by DFT, so much so that it has now become an indispensable tool for any large-scale 
calculations demanding high accuracy. The conspicuous success lies in its unique ability to strike 
a balance between quantitative accuracy, efficient computational resource in conjunction with 
readily interpretable, conceptual simplicity. This is possible because the perennial problem of 
many-body electron correlation is dealt with indirectly, but satisfactorily, through 
introduction of a fictitious \emph{non-interacting} system having \emph{same} ground state 
density of the \emph{real} system concerned. The three-dimensional, physically realizable, 
single-particle electron density (in contrast to the complicated, complex-valued, many-body wave 
function in traditional wave function-based approaches; a function of 3N space and N spin 
coordinates) takes centrestage. A major crux of the problem is now transferred in finding the 
elusive, yet-to-be-found, all-important exchange-correlation (XC) density functional, whose exact 
form still remains unknown; hence must be approximated. For small molecules while DFT can be seen
as an \emph{alternative} to the conventional \emph{ab initio} methods such as full configuration 
interaction (CI), many-body perturbation theory or coupled cluster approaches, for electronic structure 
calculations of larger complicated molecules (having chemical, physical, biological interest), DFT 
has an edge over these methods. So far this appears to be the most practical route in terms of 
accuracy and efficiency. 

This chapter focuses on an in-depth account of work done by me in past three years or so on 
stationary atomic and molecular ground states, where relativistic effects are chemically 
insignificant; therefore allowing us to use the nonrelativistic time-independent Kohn-Sham (KS) equation. 
Adiabatic partitioning of the system's wave functions into electronic and nuclear portions are 
tacitly assumed within Born-Oppenheimer approximation; thus electronic part can be separated 
from nuclear counterpart. We propose an alternative to the most commonly used and popular 
approach of the so-called atom-centered grid (ACG), by using a much simpler cartesian coordinate grid 
(CCG) to obtain electronic structure of atoms, molecules. The pertinent KS equation 
is solved by linearly expanding the desired molecular orbitals (MO) in terms of a set of known 
localized, atom-centered basis set within an LCAO-MO approximation. Analytical, one-electron 
\emph{ab-initio} pseudopotentials (expressed in terms of a sum of Gaussian functions) represent core 
electrons, whereas energy-optimized truncated Gaussian basis sets are used for valence electrons. It is 
demonstrated that our obtained self-consistent eigenvalues, eigenfunctions (and other properties
derived from them) are highly accurate, reliable and produce practically identical results as those 
obtained from their ACG counterparts. Illustrative results are given for a number of local 
and global properties (such as energy components, potential energy surface, etc.) to assess, 
establish the efficacy and relevance of our approach for a modest set of chemical species using 
both local, non-local XC density functionals.

\section{The methodology}
\subsection{A brief review of DFT}
This subsection gives an overview of DFT to the extent of summarizing only the essential ingredients 
as needed for our future discussion. By no means it is a review, because it is not needed as already 
there exists an extensively 
large number of excellent articles covering numerous aspects of the theory; moreover it is impossible
to keep pace with the innumerable amount of work that has been published on the subject. 
Application area are diverse starting from atoms to condensed matter to computational material 
science. For more detailed exposition, references \cite{march87,parr89, 
jones89,chong95,seminario96,joulbert98,dobson98,koch01,parr02,fiolhais03,gidopoulos03} may be 
consulted.

The important idea of expressing part or all of the energy of a many-electron system as functional 
of single-particle electron density has its root in seminal works of Thomas-Fermi-Dirac as early as in 
1927. The kinetic energy and exchange energy can be approximated as explicit functionals of electron 
density; idealized as non-interacting electrons in a homogeneous gas with density equal to the local 
density at any given point,
\begin{equation}
E_{TFD}[\rho(\rvec)]= C_F \int \rho(\rvec)^{5/3} \ \mathrm{d}\rvec + 
\int \rho(\rvec) v(\rvec) \ \mathrm{d} \rvec - C_x \int \rho(\rvec)^{4/3} \ \mathrm{d}\rvec
+ \frac{1}{2} \int \int \frac{ \rho(\rvec) \rho (\rvec')}{|\rvec - \rvec'|} \ \mathrm{d}\rvec
\mathrm{d} \rvec',
\end{equation}
where $C_F= \frac{3}{10}(3\pi^2)^{2/3}$, $C_x=\frac{3}{4} (\frac{3}{\pi})^{1/3}$ correspond to the
local approximations to kinetic energy and exchange energy respectively in first and third 
terms in the right hand side. The second and last terms refer to nuclear-electron attraction and 
classical electrostatic Hartree repulsion respectively. While the stunning simplicity of replacing 
complicated many-body Schr\"{o}dinger equation by a \emph{single} equation in terms of electron density 
\emph{alone}
is conspicuously appealing, underlying approximations are too crude and inaccurate to 
offer any practical applications in modern quantum chemistry; also the essential physics and 
chemistry is missing (e.g., it fails to explain shell structure in atoms and molecular binding 
\cite{teller62}). 

Insurmountable difficulties are encountered to approach such a problem beyond gross levels of 
Thomas-Fermi 
model. Consequently DFT was lost into oblivion until 1964, when it got a rebirth in the landmark 
works of Hohenberg-Kohn-Sham~\cite{hohenberg64, kohn65}. It was placed on a rigorous theoretical 
foundation thenceforth. Following an astonishingly simple proof, Hohenberg-Kohn theorem states that 
for a many-electron system, external interacting potential $v(\rvec)$ is completely determined 
(within a trivial constant) by $\rho(\rvec)$. Since the Hamiltonian is thus fully defined 
uniquely, it immediately follows that ground-state wave function and all other properties of such a  
system can be obtained from $\rho(\rvec)$ alone. Further, they proved the existence of a 
\emph{universal functional}, $F_{HK}[\rho]$, for \emph{any} valid well-behaved external potential, 
whose global minimum leads to exact ground-state energy and the density which minimizes 
this functional corresponds to exact density,
\begin{eqnarray}
E_v [\rho] & \equiv & F_{HK}[\rho] + \int v(\rvec) \rho (\rvec) \mathrm{d}\rvec \ \ \ge 
E_v [\rho_0] \\
F_{HK}[\rho] & = & \langle \Psi | T+V_{ee} | \Psi \rangle. \nonumber 
\end{eqnarray}
Here $\Psi$ denotes ground-state wave function associated with $\rho$, $E_v[\rho_0]$ is 
ground-state energy of the Hamiltonian with external potential $v(\rvec)$, $\rho_0$ is its 
ground-state density, $V_{ee}$ denotes two-particle interaction energies including both 
classical and non-classical effects.

While Hohenberg-Kohn theorem physically justifies earlier works of Thomas-Fermi and others 
to employ $\rho(\rvec)$ as a central variable to describe a many-electron system, it is a 
\emph{proof of existence} only. All it asserts is that there is a unique mapping between 
ground-state density and energy, in principle; but it keeps mute on furnishing any information on 
the construction of such a functional. Although $\rho(\rvec)$ is sufficient, the relation is subtle,
intricate, and how to extract any set of general properties from it, remains unknown. Another important 
disconcerting feature is that minimization of $E_v[\rho]$ is, in general, a tough numerical task.
Actual calculations are still as hard as before; there is absolutely no clue on what kind of 
approximations to be used for unknown functionals. In effect, no visible progress could be 
discerned for realistic calculations (no simplification over MO theory), and hence it is of not much 
help in quantum chemistry, because the final step still constitutes solution of Schr\"{o}dinger 
equation, which is prohibitively difficult. Another awkward dilemma arises in density 
variation principle, which holds true only for \emph{exact} functionals. Thus, in contrast to 
conventional wave-function-based approaches, where, e.g., Hartree-Fock (HF) or Configuration 
interaction wave functions are strictly variational, within the rubric of Hohenberg-Kohn DFT, energy 
delivered by a trial functional has absolutely no physical meaning whatsoever.

In a breakthrough work, Kohn and Sham, in order to alleviate these outstanding difficulties, 
introduced the concept of a hypothetical \emph{non-interacting} reference system having   
\emph{same} overall ground-state density as our real interacting system. This Hamiltonian can now be 
expressed as sum a of one-electron operators, has eigenfunctions that are Slater determinants of 
individual one-electron eigenfunctions and has eigenvalues that are sum of one-electron 
eigenvalues. This realization of a non-interacting system, built from a set of one-electron 
functions (orbitals) came from an observation that conventional orbital-based approaches such as HF 
or so fares better in this regard. Thus a major chunk of actual kinetic energy can now be recovered 
fairly accurately as a sum of individual electronic kinetic energies. The residual, often small, 
contribution ($T_c$) is merged with the non-classical contributions to electrostatic repulsion, 
whose \emph{exact} form is also unknown,
\begin{eqnarray}
F[\rho] & = & T_s[\rho] + J[\rho] + E_{xc} [\rho] \\ 
E_{xc}[\rho] & = & (T[\rho] - T_s[\rho]) + (V_{ee}[\rho] - J[\rho]) = 
T_c[\rho] + (V_{ee}[\rho] - J[\rho]). \nonumber 
\end{eqnarray}
Here the subscript in $T_s$ denotes independent-particle kinetic energy of a non-interacting 
system, $V_{ee}[\rho]$ signifies two-particle repulsion energy while $J[\rho]$ the classical part of 
$V_{ee}[\rho]$, $E_{xc}[\rho]$ refers to XC energy. Kohn-Sham (KS) orbital equations, in their 
canonical form, are written as follows (henceforth atomic units implied, unless otherwise mentioned),
\begin{equation}
\left[ -\frac{1}{2} \nabla^2 + v_{eff} (\rvec) \right] \psi_i (\rvec) = \epsilon_i \psi_i (\rvec)
\end{equation}
\begin{equation}
v_{eff}(\rvec) = v(\rvec) + \int \frac{\rho(\rvec')}{|\rvec - \rvec'|} \ \mathbf{d} \rvec' 
+ v_{xc}(\rvec). \nonumber
\end{equation}
Here $v_{eff}(\rvec)$ is the effective potential and $v(\rvec)$ gives the external potential 
due to nuclei or any other external field (assumed to be spin-independent). It is noteworthy that 
although working equations of KS DFT are very similar to those of HF theory, the most profound 
difference between them lies in that KS theory incorporates the intriguing, delicate electron 
correlation effects rigorously; thus, it is, in principle, capable of yielding \emph{exact} 
Schr\"{o}dinger energy, which is certainly not the case for latter. 

\subsection{LCAO-MO ansatz of DFT}
As already hinted, minimization of the explicit functional is not normally the most efficient and
recommended path towards an actual working DFT. A far more attractive practical route is through 
KS equation, that owes its success and 
popularity, interestingly, partly due to the fact that it does not work solely in terms of particle 
density, but rather brings back an orbital picture into the problem. Hence, in essence, formally 
it seems to appear like a single-particle theory. Nevertheless many-body effects are incorporated 
\emph{in principle}, exactly.

In some straightforward implementation, the so-called real-space method \cite{beck00}, instead of 
expanding wave functions in a suitably chosen, predetermined basis set, latter are normally sampled 
in a real-space grid, usually through either of the 
following three representations, such as finite difference (FD), finite element or wavelets. In all these 
cases, however, relevant discrete differential equations produce highly structured, banded, 
sparse matrices. This representation has advantages that the potential operator is diagonal in 
coordinate space; also Laplacian is nearly local (that makes them good candidates for linear-scaling 
approaches) and these are easily amenable to domain-decomposition parallel implementation. Moreover, 
Hartree potential can be found using highly optimized FFTs or real-space multigrid algorithms. 
Some of the earliest successful works along this direction were by \cite{laaksonen85,becke89,
white89}, where a basis-set-free, fully numerically converged, FD approach 
was adopted for solution of self-consistent eigenvalue problems encountered in atomic, molecular 
cases, with reasonably good accuracy. Thereafter, a polyatomic numerical integration scheme was devised, 
whereby 
the actual physical domain was partitioned into a collection of single-centre components having 
radial grids centered at each nucleus. While these results rekindled hope of employing grid-based 
methods for full-scale atomic and molecular studies in quantum chemistry, they lacked the highly 
desirable scalability and efficiency requirement essential for any large-scale calculation. Later, 
instead of 
above mentioned atom-centered grids, high-order real-space pseudopotential method was used for 
relatively larger systems in uniform cartesian coordinates \cite{chelikowsky94a,chelikowsky94b}. In 
a uniform orthogonal 3D mesh containing grid points ($x_i, y_j, z_k$), $m$th order expansion of 
Laplacian operator, within an FD approximation, can be expressed as follows ($h$ denotes grid 
spacing and $m$ is a positive integer),
\begin{equation}
\left[ \frac{\partial^2 \psi}{\partial x^2} \right]_{x_i, y_j, z_k} = \sum_{-m}^{m}
C_m \psi(x_i+mh, y_j, z_k) +O (h^{2m+2}).
\end{equation}
As it is known, this approximation is valid in the limit of $h \rightarrow 0$ (i.e., finer 
grid structure). However, our common experience suggests that to obtain physically meaningful and 
sufficient accuracy, orders higher than second are most often necessary. FD methods have been applied 
to a number of interesting \emph{ab initio} self-consistent problems including clusters and other 
finite systems \cite{chelikowsky94a,chelikowsky94b,modine97,kim99,lee00}. Standard iterative 
processes are noticeably less efficient on finer messes, however, and multigrid methods have been 
proposed to accelerate the self-consistent iteration procedure \cite{briggs95,briggs96}. They have 
been used in conjunction with adaptive grids to enhance the resolution \cite{gygi95,modine97}. 
Numerical convergence is controlled by only a few parameters like grid spacing, domain size, order of 
representation, etc.  

Some of the major limitations of this method are that they are (i) non-variational and (ii) dimension
of Hamiltonian matrix is unmanageably large. The parallel, \emph{basis-set} approach, however, 
dominates quantum chemistry community and a plethora of physical, chemical, biological 
applications have been made to study energetics, dynamics, reaction mechanisms, etc., encompassing 
an astoundingly wide range of interesting systems. This is advantageous since it gives an 
opportunity to exploit the enormous advances made in previously developed techniques within the 
context of basis-set solutions of wave function-based methodologies (such as HF, for 
example). The so-called linear combination of atomic orbitals (LCAO) 
ansatz is by far the most practical, popular and convenient computational route for iterative 
solution of molecular KS equation. Denoting the one-electron KS operator in parentheses of 
Eq.~(4) by $\hat{f}^{KS}$, one can cast KS equation in following operator form, 
\begin{equation}
\hat{f}^{KS} \psi_i = \epsilon_i \psi_i.
\end{equation}
The above operator differs from Fock operator $\hat{f}^{HF}$ used in connection to HF theory 
in that former includes non-classical many-body exchange-correlation effects, $v_{xc}$ 
exactly (as a functional derivative with respect to charge density, $v_{xc}[\rho]= 
\delta E_{xc}[\rho] / \delta \rho$), whereas latter does not account for any correlation effects. 
This represents a fairly complicated system of coupled integro-differential equation (kinetic energy 
is given by a differential operator, whereas the Coulomb contribution by an integral operator), 
whose solution yields the desired KS MOs $\{\psi_i\}$. Numerical procedures for solution of this 
equation are much too demanding. Computationally the most efficient way is to linearly expand 
unknown KS MOs in terms of a K known basis functions $\{\phi_{\mu}, \mu=1,2,\cdots,K\}$, 
\begin{equation}
\psi_i = \sum_{\mu=i}^{K} C_{\mu i} \phi_{\mu}, \ \ \ i=1,2,\cdots , K
\end{equation}
in a manner analogous to 
the LCAO-MO scheme employed in Roothaan-Hartree-Fock method. For a complete set $\{\phi_{\mu}\}$ 
with $K=\infty$, above expansion is exact and any complete set could be chosen, in principle. 
However, for realistic computational purposes, one is invariably restricted to a finite set; thus it 
is of paramount importance to choose functions $\{\phi_{\mu} \}$ such that the approximate expansion 
reproduces KS orbitals as accurately as possible. Now, inserting Eq.~(8) in (7), multiplying the 
resulting equation 
from left with arbitrary basis function $\phi_{\mu}$, integrating over space, followed by some 
algebraic manipulation, leads to, in close analogy to HF case, following compact matrix equation, 
\begin{equation}
\mathbf{F}^{KS} \mathbf{C = SC} \epsilon.
\end{equation}
Here, $\mathbf{S}$ and $\mathbf{F}$ denote $K \times K$ real, symmetric overlap and total KS 
matrices respectively; eigenvector matrix $\mathbf{C}$ contains unknown expansion coefficients 
$C_{\mu i}$ whereas orbital energies $\epsilon_i$ are embedded in the diagonal matrix 
$\mathbf{\epsilon}$. Note that through the introduction of a basis set, original problem of 
optimizing a complicated, highly nonlinear integro-differential equation is now transformed into a 
linear one, which could be easily solved efficiently, accurately by using standard techniques of linear 
algebra. Individual elements of KS matrix are given as, 
\begin{eqnarray}
F_{\mu \nu}^{KS} & = & \int \phi_{\mu} (\rvec) \left[ h^{\mathrm{core}}+v_{HXC}(\rvec) \right] \phi_{\nu} 
(\rvec)
\mathrm{d} \rvec = H_{\mu \nu}^{\mathrm{core}} + \langle \phi_{\mu}(\rvec) | v_{HXC} | \phi_{\nu} 
(\rvec)
\rangle \\
 & = & H_{\mu \nu}^{\mathrm{core}} + J_{\mu \nu} +V_{\mu \nu}^{XC},  \nonumber 
\end{eqnarray}
where $H_{\mu \nu}^{\mathrm{core}}$ denotes bare-nucleus Hamiltonian matrix (including 
kinetic energy plus nuclear-electron attraction) and accounts for one-electron energies; 
$v_{HXC} (\rvec)$ determines all two-electron potentials including classical Coulomb repulsion as well as 
the non-classical XC potential. For certain choices of $\{\phi_{\mu}\}$ (such as Gaussian bases), 
one-electron matrix elements can be fairly easily computed analytically using well-tested 
algorithms. $J_{\mu \nu}$ signifies the matrix elements corresponding to classical Hartree 
repulsion term, which is expressed in terms of electron density which takes the following form 
within LCAO approximation,
\begin{equation}
\rho(\rvec) = \sum_{i=1}^{N} \sum_{\mu=1}^{K} \sum_{\nu=1}^{K} C_{\mu i} C_{\nu i} 
\phi_{\mu}(\rvec) \phi_{\nu}(\rvec).
\end{equation} 
The remaining term, $V_{\mu \nu}^{XC}$ represents matrix element of XC contribution and 
constitutes the most difficult step of the whole process.

\subsection{Basis set and Pseudopotential}
Design and choice of an appropriate basis set for a particular problem is a very crucial step in molecular 
calculation. Stunningly large volumes of work have been done towards their construction as well as 
effects on various physicochemical properties. Broadly speaking, there are two major considerations 
besides accuracy: (a) reduction of number of functions in the expansion (b) ease and efficiency of the 
computation of relevant integrals. At the onset, it is anticipated that basis set requirement for 
wave function-based and density-based approaches should be quite different. In the former, MOs generate
approximate wave 
function, while in latter scenario, orbitals enter in to the picture indirectly as a tool to 
generate charge density. Ever since the inception of LCAO-MO procedures in quantum chemistry, a 
significantly large number of versatile, elegant, flexible basis sets have been developed in context 
of molecular calculation (for a variety of situations such as HF method, correlated post-HF methods, ground 
and excited states, weakly interacting systems, diffused systems such as anions etc). This remains to be 
a very tricky and delicate
problem, at best, because as yet, there is no universal molecular basis set applicable for all methods or
chemical systems of interest. Many decent reviews are available \cite{szabo96,helgaker00,
cramer04,jensen07} on this topic and thus our current disposition pertains only to the point as 
needed for our future discussion.

Some commonly used basis functions are: for periodic systems (e.g., solids), plane wave whereas for 
non-periodic systems (e.g., molecules, clusters), atom-centered localized basis sets such as Slater 
type orbitals (STO), Gaussian type orbitals (GTO), numerical radial functions, linear muffin-tin 
orbitals, delta functions, etc. Combination of basis sets have also been devised; e.g., in a 
Gaussian and augmented plane wave approach \cite{lippert99,krack00}, KS MOs and electron charge 
densities are expanded in Gaussian and an auxiliary augmented plane wave basis sets respectively. 
However, amongst all these, GTOs have gained maximum popularity over others and have remained the most
preferred option for long time (also used in this work) 
chiefly due to their computational advantages offered for various multicenter one- and two-electron 
integrals involved (efficient analytical algorithms can be employed for these). STOs, on the other 
hand, although, were, at first used as a natural choice, for they correspond to atomic orbitals 
of H atom (and hence offer a better qualitative description of molecular orbitals), were later 
disfavored because of difficulty in evaluating the aforementioned integrals (no analytical routes 
are known and recourse must be taken to numerical techniques). Cartesian GTOs are typically written 
as,
\begin{equation}
\phi(\zeta,n_x,n_y,n_z; x,y,z) = N x^{n_x} y^{n_y} z^{n_z} \ e^{-\zeta r^2}.
\end{equation}
In the above equation, $N$ denotes normalization constant, while $\zeta$, the orbital exponent, 
characterizes compactness (large $\zeta$) or diffuseness (small $\zeta$) of the function. Functions with 
$\lambda=n_x+n_y+n_z=0,1,2,\cdots,$ are referred to as $s,p,d,\cdots,$ respectively, with $\lambda$ 
being closely related to total angular momentum quantum number. It is easy to approximate the shape of 
an STO function by summing up a number of GTOs with different exponents and coefficients. However, 
usually one needs about three times as many GTOs than STO functions to achieve a certain desired 
accuracy. This dilemma is resolved by employing the so-called \emph{contracted} GTOs consisting of 
a fixed linear combination (i.e., same coefficients and exponents) of the \emph{primitive} 
functions, $\phi_p$, 
\begin{equation}
\phi_{\mu}^{\mathrm{CGF}} (\rvec- \mathrm{\mathbf{R_A}}) = \sum_{p=1}^{L} d_{p \mu} 
\phi_p(\zeta_{p\mu}, \rvec- \mathrm{\mathbf{R_A}}).
\end{equation}
Here $L$ corresponds to the length of contraction, $d_{p \mu}$ is contraction coefficient, ``p" 
symbolizes primitive functions while ``CGF" stands for contracted Gaussian functions. Gaussian 
primitives are often optimized from atomic calculations (ideally HF or CI, etc.) variationally 
until lowest total energy of the atom achieved; there are also cases where they are explicitly 
optimized through KS scheme using XC functionals of homogeneous electron gas. However, interestingly,
basis sets originally designed for wave function-based methodologies apparently work fairly well, as 
fortunately it turns out that, for most of the common 
important properties (like energy, equilibrium geometry, etc.), the results are fairly insensitive 
with respect to the way exponents and coefficients in a basis set have been determined. This gives us a 
pleasant opportunity for it validates the use of basis sets, primarily designed for wave function 
based methods, within the rubric of DFT, with reasonable confidence.  

Now it is well known that many-electron HF or KS equation could be much simplified by dividing  
electrons into core and valence categories. Inner-shell electrons, being strongly bound to the 
nucleus (forming the so-called ``inert" core and thus to a good approximation, retain an atomic-like 
configuration), play less significant role in chemical binding and most of the chemical properties 
can almost completely be accounted for by taking care of valence shells only. In essence, this 
facilitates to replace the strong Coulomb potential of nucleus and tightly bound core 
electrons by an effective potential (smoother, non-local) acting on the valence electrons. Under these
circumstances, our desired KS equation in presence of pseudopotential could be rewritten as, 
\begin{equation}
\left[ -\frac{1}{2} \nabla^2 +v^p_{ion}(\rvec) + v_H[\rho](\rvec) + v_{xc}[\rho](\rvec) \right] 
\psi_i (\rvec) = \epsilon_i \psi_i(\rvec),
\end{equation}
where $v^p_{ion}$ denotes ionic pseudopotential for the system,
\begin{equation}
v^p_{ion}= \sum_{R_a} v^p_{ion,a}(\rvec-\mathbf{R_a}).
\end{equation}
Here, $v^p_{ion,a}$ is the ion-core pseudopotential associated with atom A, situated at $R_a$;
$v_H[\rho](\rvec)$ describes the classical electrostatic interaction among valence electrons, while 
$v_{xc}[\rho](\rvec)$ represents non-classical part of the many-electron Hamiltonian.

A host of different pseudopotentials have been developed by many workers over the years, such as 
empirical, \emph{ab initio}, 
norm-conserving, ultrasoft, etc., \cite{hamann79,bachelet82,vanderbilt90,troullier91,payne92,
hartwigsen98, martin04}. This work uses the \emph{ab initio} effective core potentials developed by 
\cite{wadt85,hay85}, where total potential is expressed in terms of projection operators,
$P_l=|l \rangle \langle l|$, as, 
\begin{equation}
U(r)=U_L(r) + \sum_{l=0}^{L-1} \left [ U_l(r)-U_L(r) \right] P_l.
\end{equation} 
In this equation, $U_l$ identifies numerical effective core potentials for each $l$. For computational 
conveniences, an analytic form for $U(r)$ is obtained by fitting, 
\begin{eqnarray}
r^2 \left[ U_l(r) -U_L(r) \right], &   &  l= 0,1,\cdots,L-1, \\
r^2 \left[ U_l(r) -N_c/r \right], &   &  l= L, \nonumber 
\end{eqnarray}
to a Gaussian of the form $\sum_k d_k r^{n_k} e^{-\zeta_kr^2}$. Here, $L$ signifies the lowest angular 
momentum not represented in core, $N_c$ is number of core electrons and $n_k=0,1,2$.

\subsection{Computational considerations}
In this subsection we discuss various practical problems encountered in dealing with the solution 
of a molecular KS equation within the basis-set framework. Before we get into mind-boggling details, 
it is noteworthy to mention, on the outset, that formally HF calculation scales as $N^4$, while KS 
calculations do so no worse than $N^3$, where $N$ stands for number of basis functions. Clearly, this 
is better than HF by a factor of $N$ and significantly better than other traditional correlated methods. 
This $N^4$ originates from the total number of two-electron repulsion integrals (proportional to $N^4$)
in the HF case. However later it has been argued that well-implemented HF or KS programs for large 
systems, in effect, scale as $N^2$ only, if one takes consideration of the fact that a vast majority 
of these two-electron integrals are essentially zero 
due to negligible overlap among basis functions involved.

A very widely used LCAO-MO-based DFT procedure \cite{andzelm92} introduces an \emph{auxiliary} basis 
set (in addition to the one used for MO expansion) to fit (often some variation of least square 
technique) some of the computationally intensive terms to reduce the 
integral evaluation overhead significantly, making this an $N^3$ process. Some of the earliest notable
works in this direction \cite{sambe75,dunlap79,dunlap79a} proposed following expansion for electron 
density and XC potential in terms of auxiliary basis sets, $f_i$ and $g_j$ respectively as,
\begin{eqnarray}
\rho(\rvec) & \approx & \tilde{\rho}(\rvec)   =  \sum_i a_i f_i(\rvec) \\
v_{xc}(\rvec) & \approx & \tilde{v}_{xc}(\rvec)   =  \sum_j b_j g_j(\rvec), \nonumber
\end{eqnarray}
where tildes identify fitted quantities and $\{a_i\}, \{b_j\}$ are the fitting coefficients. The latter 
are determined by minimization of either the following straightforward function,
\begin{equation}
Z=\int [\rho(\rvec)-\tilde{\rho}(\rvec)]^2 \mathrm{d}\rvec,
\end{equation}
or Coulomb self-repulsion of residual density. Both, of course, are subject to the constraint that 
normalizaion of this fitted density gives total number of electrons. Originally, this technique was 
first suggested in the context of STOs \cite{baerends73} and later extended to GTOs \cite{sambe75}. 
Typically XC potentials are calculated in real-space; corresponding matrix elements are evaluated 
by some analytical means.

While the above fitting procedure has witnessed tremendous success in explaining many chemical phenomena
within DFT, it suffers from some 
serious discomfitures. Firstly, there are many different flavors of fitting techniques (variational 
as well as non-variational) available which remains a primary source of inconsistency among various 
implementations; also constraints imposed on density and XC fitting are different, thus aggravating 
the problem. Second, number of electrons is not automatically conserved in fitted density; 
neither does it reproduce true multipole moments exactly. Another important shortcoming of this 
scheme is that this greatly complicates analytic derivative theories. Last, but not the least, it is 
worth mentioning that, the historical motivation for development of fitting method arose primarily 
due to the lack 
of efficient, good-quality integral methods at that time. However thereafter, a plethora of work was 
reported by various workers on multi-center molecular integrals (see \cite{gill94} for a review); thus 
once seemingly daunting task of 
integration evaluation, although even now poses a major efficiency issue, has found many attractive, 
elegant, high-quality, manageable and satisfactory approximations. As expected, however, the appeal of 
fitting method somehow dwindled with time, especially after the emergence of numerous highly accurate 
efficient quadrature schemes.  

Solution of our desired KS equation always involves mathematically non-trivial integrals. Unlike 
the exchange integrals of HF theory which can be analytically calculated for Gaussians basis sets, 
unfortunately, XC functionals (by
virtue of their complicated approximate algebraic forms) in DFT require computation of integrals, 
for which, as yet, there is no easy analytic route available and almost invariably must only be done by 
numerical methods. One of the most successful, popular and generic schemes (so-called atom-centered 
grid (ACG)), 
is due to \cite{becke88}. In this case, a 3D molecular integrand is partitioned into single-center 
discrete, overlapping atomic ``cells". For an arbitrary integrand $F(\rvec)$, this decomposition 
gives the value of integral $I$, as,
\begin{equation}
I=\int F(\rvec) \mathrm{d} \rvec = \sum_A^M I_A.
\end{equation} 
Here, the sum runs over all $M$ nuclei present in the system, and $I_A=\int F_A(\rvec) \mathrm{d} 
\rvec$ denotes single-center atomic contributions in it. Atomic integrands, $F_A$ are chosen such that 
when summed over all the nuclei $M$, they should return our original function,
\begin{equation}
\sum_A^M F_A (\rvec) = F(\rvec). 
\end{equation}
The individual components $F_A(\rvec)$ are finally constructed from original integrand through some
sufficiently well-behaved weight functions, $F_A (\rvec)= w_A(\rvec) F(\rvec)$. Thus each of the
atomic subintegrations can be carried out using standard mono-centric numerical techniques. Once 
$F_A$s are determined, the corresponding integrals $I_A$ are subsequently computed on grid as 
follows (assuming polar coordinates, which is the commonest choice in literature),
\begin{equation}
I_A= \int_{0}^{\infty} \int_{0}^{\pi} \int_{0}^{2\pi} F_A(r,\theta,\phi) r^2 \sin \theta 
\ \mathrm{d} r \ \mathrm{d}\theta \ \mathrm{d}\phi \approx \sum_{p}^{P} w_p^{\mathrm{rad}}
\sum_{q}^{Q} w_q^{\mathrm{ang}} \ F_A(r_p, \theta_q, \phi_q),
\end{equation}
where $w_p^{\mathrm{rad}}$, $w_q^{\mathrm{ang}}$ signify radial, angular weights respectively
with $P$, $Q$ points (total number of points being $P \times Q$). Usually 
angular part is not further split into separate $\theta$, $\phi$ contributions, as surface 
integrations on a sphere can be done numerically quite easily accurately by the help of available highly 
efficient algorithms.

Judicious choice and determination of proper weight functions involve certain amount of mathematical 
subtlety and requires both scientific as well as artistic skill. Detailed account on this and various 
other related issues can be found in 
following references \cite{becke88,murray93,treutler95}. A number of elegant and clever prescriptions 
have been 
proposed for both radial as well as angular integrations, depending on the particular quadrature 
scheme and mapping functions. Some of the most promising choices for former are: Gauss-Chebyshev 
quadrature of second kind, Chebyshev quadrature of first and second kind, Gaussian quadrature, 
Euler-MacLaurin formula, numerical quadrature \cite{murray93,treutler95,mura96,gill93,lindh01}, etc. 
For angular integrations, there is some sort of consensus that octahedral Lebedev grid 
\cite{lebedev75,lebedev76,lebedev92} is possibly the most efficient and satisfactory. Numerically 3D 
integration of molecular 
integrands have also been achieved based on a division of space and applying product Gauss rule for 
subsequent integration of resulting regions \cite{boerrigter88}. A variational integration 
mesh \cite{pederson90}, depending on the position of individual atoms in three different regions of
space, \emph{viz.,} atomic spheres, excluded cubic region and interstitial parallelepiped, has been 
reported as well. In another development \cite{krack98}, an automatic numerical integrator capable 
of generating adaptive molecular grid depending on size and shell structure of a given basis set,
was also suggested.

The above observation clearly points out that LCAO-MO-based KS DFT calculations of molecules have 
largely been done through ACG. Very little attempt has been made to test other grids in this 
occasion. Of particular interest is the much simpler cartesian coordinate grid (CCG). To our knowledge, 
virtually 
no results are available to judge their performance. In a Fourier Transform Coulomb 
method \cite{fustimolnar02}, molecular integrations have been approached by separating Gaussian
shell pairs into ``smooth" and ``sharp" categories. Of late, a multiresolution technique has also
been proposed, where a connection between CCG and ACG is made via a divided-difference polynomial 
interpolation scheme, which virtually translates the electron density and gradients from former to 
latter \cite{kong06}. Our primary objective here is to report the development of a new, successful DFT
methodology \cite{roy08,roy08a} which utilizes CCG. This has been shown to produce highly accurate 
dependable  results for small to medium sized atoms/molecules using Gaussian basis set.
For a series of atoms and molecules, results offered were almost identical to those obtained from 
other existing reliable calculations available (especially those involving ACG). Atom-centered basis 
functions, electron densities, MOs, as well as Hartree and XC potentials all are directly built up on a 
real uniform 3D Cartesian grid simulating a cubic box (although nonuniform grids could be used for 
this purpose, and we are currently engaged into it),
\begin{equation}
r_i=r_0+(i-1)h_r, \ \ \ i=1,2,\cdots,N_r; \ \ \ \mathrm{for} \ r \in \{x,y,z\},
\end{equation}
where $h_r, N_r$ signify the grid spacing and total number of grid points respectively 
($r_0=-N_r h_r/2$). Electron density in this grid is then simply given by, 
\begin{equation}
\rho(\rvec) = \sum_{i}^N |\psi_i(\rvec)|^2 = \sum_{\mu=1}^K \sum_{\nu=1}^K P_{\mu \nu} \ 
\phi_{\mu} (\rvec) \ \phi_{\nu} (\rvec). 
\end{equation} 
Here $P_{\mu \nu}$ represents an element of the density matrix.

Now we turn our focus on to a discussion on calculation of classical electrostatic repulsion potential 
in the grid. For finite systems, possibly the simplest and crudest means to compute $v_H(\rvec)$ 
consists in a direct numerical integration, which, in many occasions show sluggish performance, and
in general, remains feasible only for relatively smaller systems. However the most successful and
favorite route is through a solution of corresponding Poisson equation. An alternate accurate
and efficient technique utilizes conventional Fourier convolution method and several of its variants
\cite{martyna99,minary02}; this has been immensely successful in the context of molecular modeling 
in recent years, 
\begin{eqnarray}
\rho(\mathrm{\mathbf{k}}) & = & \mathrm{FFT} \{ \rho(\rvec) \}  \\
v_H(\rvec) & = & \mathrm{FFT}^{-1}\{ v_H^c(\mathrm{\mathbf{k}}) \rho(\mathrm{\mathbf{k}}). \nonumber
\end{eqnarray}
Here $\rho(\mathrm{\mathbf{k}})$ and $v_H^c(\mathrm{\mathbf{k}})$ represent Fourier integrals of
density and Coulomb interaction kernel, respectively, in the grid. The former is customarily obtained 
from a discrete Fourier transform of its real-space value by standard FFT quite easily. Evaluation of 
the latter, however, is a non-trivial task because of the presence of singularity in real space and 
demands caution.
In our current communication, this is overcome by applying a decomposition of the kernel into long- and 
short-range interactions, reminiscent of the commonly used Ewald summation technique in condensed
matter physics, 
\begin{equation}
v_H^c(\rvec) = \frac{\mathrm{erf}(\alpha r)} {r} + \frac{\mathrm{erfc} (\alpha r)}{r} \equiv
v^c_{H_{\mathrm{long}}} (\rvec) + v^c_{H_{\mathrm{short}}} (\rvec),
\end{equation}
where erf(x) and erfc(x) correspond to the error function and its complements respectively. Short-range 
Fourier integral can be calculated analytically whereas the long-range contribution can be obtained 
directly from FFT of real-space values. There are several other routes as well available for classical 
repulsion as needed in the large-scale electronic structure within KS DFT framework. For example, in 
the fast 
multipole method, near-field contributions are handled explicitly, while the far-field is treated 
through a clustering of spatial cells, whereas the field itself is expressed by a multipole expansion. 
Another
efficient route employs a multigrid method within real-space formalism (such as finite difference or
finite element). More thorough account on this topic is covered in the review \cite{beck00}. 

Now a few words on the construction of respective one- and two-body matrix elements is in order. All 
the one-electron integrals like overlap, kinetic-energy, nuclear-attraction integral, pseudopotential 
integral as well as their respective matrix elements are completely identical to those encountered in 
standard GTO-based HF calculation; they are performed by following standard recursion algorithms
\cite{obara86,mcmurchie78,kahn76}. In absence of any analytical method, respective two-electron matrix
elements in the real-grid are computed through direct numerical integration, 
\begin{equation}
\langle \phi_{\mu} (\rvec) | v_{HXC} (\rvec) | \phi_{\nu} (\rvec) \rangle = h_x h_y h_z 
\sum_{\mathrm{grid}} \phi_{\mu} (\rvec) v_{HXC} (\rvec) \phi_{\nu} (\rvec).
\end{equation}
The matrix eigenvalue problem is accurately and efficiently solved using standard LAPACK routines 
\cite{anderson99} following usual self-consistent procedure to obtain KS eigenvalues and 
eigenfunctions, from which total energies and/or other quantities can be calculated. In all the results 
presented in next section, convergence of three quantities, \emph{viz.,} (i) potential (ii) total 
energies and (iii) eigenvalues has been monitored. Tolerance of $10^{-6}$ a.u., was employed for (ii), 
(iii), while $10^{-5}$ a.u., for (i).

\begingroup
\squeezetable
\begin{table}      
\caption{\label{tab:table1} Energy components and total number of electrons (in a.u.) 
for Cl$_2$ compared with reference values at $R=4.20$ a.u., for various grids.}
\begin{ruledtabular}
\begin{tabular} {lccccccccc}
    & \multicolumn{2}{c}{$N_r=32$}  &  \multicolumn{3}{c}{$N_r=64$}   & \multicolumn{2}{c}{$N_r=128$}  &  
\multicolumn{1}{c}{$N_r=256$}   & Ref. \cite{schmidt93}    \\
\cline{2-3} \cline{4-6} \cline{7-8} \cline{9-9} 
Set  & A & B & C & D &   E        &  F         &    G       &  H   &   \\
$h_r$                      &  0.3       &   0.4      &   0.2      & 0.3        & 0.4        & 0.1        
                           & 0.2        & 0.1        &            \\
$\langle T \rangle$        &  11.00750  & 11.17919   &  11.18733  & 11.07195   & 11.06448   & 11.18701   
                           & 11.07244   & 11.07244   & 11.07320    \\
$\langle V^{ne}_t \rangle$ & $-$83.43381& $-$83.68501& $-$83.70054& $-$83.45722& $-$83.44290& $-$83.69988
                           & $-$83.45810& $-$83.45810& $-$83.45964 \\
$\langle V^{ee}_t \rangle$ & 32.34338   & 31.22265   & 31.22885   & 31.00981   & 31.00306   & 31.22832   
                           & 31.01000   & 31.01000   &  31.01078   \\
$\langle V \rangle$        & $-$39.42376& $-$40.79570& $-$40.80503& $-$40.78074& $-$40.77317& $-$40.80489
                           & $-$40.78144& $-$40.78144&  $-$40.78219\\
$\langle E_{el} \rangle$   & $-$40.08293& $-$41.28318& $-$41.28437& $-$41.37545& $-$41.37535& $-$41.28455
                           & $-$41.37566& $-$41.37566&  $-$41.37566\\
$\langle E \rangle$        & $-$28.41626& $-$29.61651& $-$29.61770& $-$29.70878& $-$29.70868& $-$29.61789
                           & $-$29.70900& $-$29.70900&  $-$29.70899\footnotemark[1] \\
 $N$                       & 13.89834   & 13.99939   &  13.99865  & 14.00002   &  14.00003  & 13.99864   
                           &  14.00000     &  13.99999         & 13.99998  \\
\end{tabular}
\end{ruledtabular}
\footnotetext[1]{This is from grid-DFT calculation; corresponding grid-free DFT value is 
$-$29.71530 a.u.} 
\end{table}
\endgroup
\section{Results and discussion}
At first, we will examine the stability and convergence of our calculated nonrelativistic ground-state 
energies as well as individual energy components for a representative molecule, Cl$_2$ with respect to 
grid parameters (8 Sets are given), in Table I, at an internuclear distance 4.20 a.u., 
for LDA XC potential. These are to to compared with the widely used GAMESS \cite{schmidt93} quantum chemistry 
program, with same XC functional, basis set and effective core potential. The homogeneous electron-gas 
correlation of Vosko-Wilk-Nusair (VWN) \cite{vosko80} is employed in all LDA calculations. The Hay-Wadt (HW) 
valence basis set, used in this work, has the orbital split into inner and outer components (described by two 
and one primitive Gaussians respectively). Total integrated electron density, $N$ is given as well. Reference 
total energies obtained from ``grid" and ``grid-free" DFT are quoted; former using a default ``army" grade grid
with Euler-McLaurin quadratures for radial integration and Gauss-Legendre quadrature for angular 
integrations. Grid-free approach \cite{zheng93,glaesemann98}, on the other hand, utilizes a resolution of identity
to simplify molecular integrals over functionals rather than in quadrature grids. Most appealing feature
of such a method is that it enables one to bypass any error associated with a finite grid; however that
usually requires an auxiliary basis set to expand the identity itself, which also suffers from 
inherent weakness of incompleteness. First we note that, our maximum deviation in energy and $N$ with 
respect to reference values is observed in Set A, mainly because our box length is insufficient to
capture all important interactions present. At Set B, with an increase in spacing, box gets
enlarged and all these quantities get significantly bettered than in Set A. Notice that Sets C and F both 
produce very similar results for all quantities including $N$, which is expected intuitively, as they both 
cover same box length. As Set B also corresponds to same box size as the above two Sets, results are
again comparable; however, component energies and $N$ show slight deviations. Sets D, E give 
all energy quantities as well as $N$ in very good agreement with reference values. For further 
verification, some calculations are also performed in relatively larger finer grids F, G, H. From 
above discussion it is easily concluded that Sets D, E, G, H are our four best results, while Sets D, E are 
sufficiently accurate for all practical purposes. More detailed discussion on this (for Cl$_2$, HCl) could be found in \cite{roy08}.

\begin{figure}
\begin{minipage}[c]{0.40\textwidth}
\centering
\includegraphics[scale=0.45]{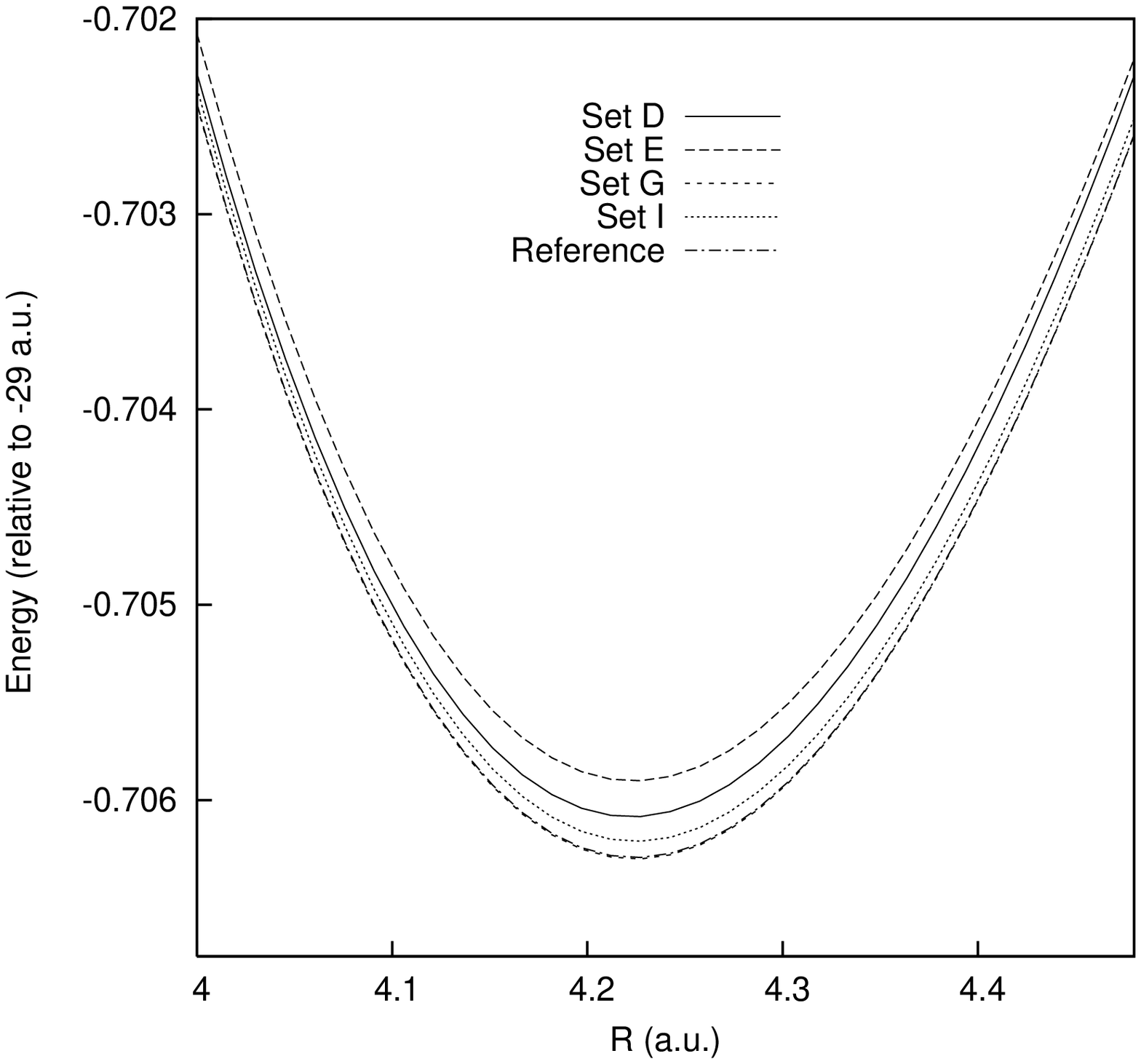}
\end{minipage}%
\hspace{0.5in}
\begin{minipage}[c]{0.40\textwidth}
\centering
\includegraphics[scale=0.45]{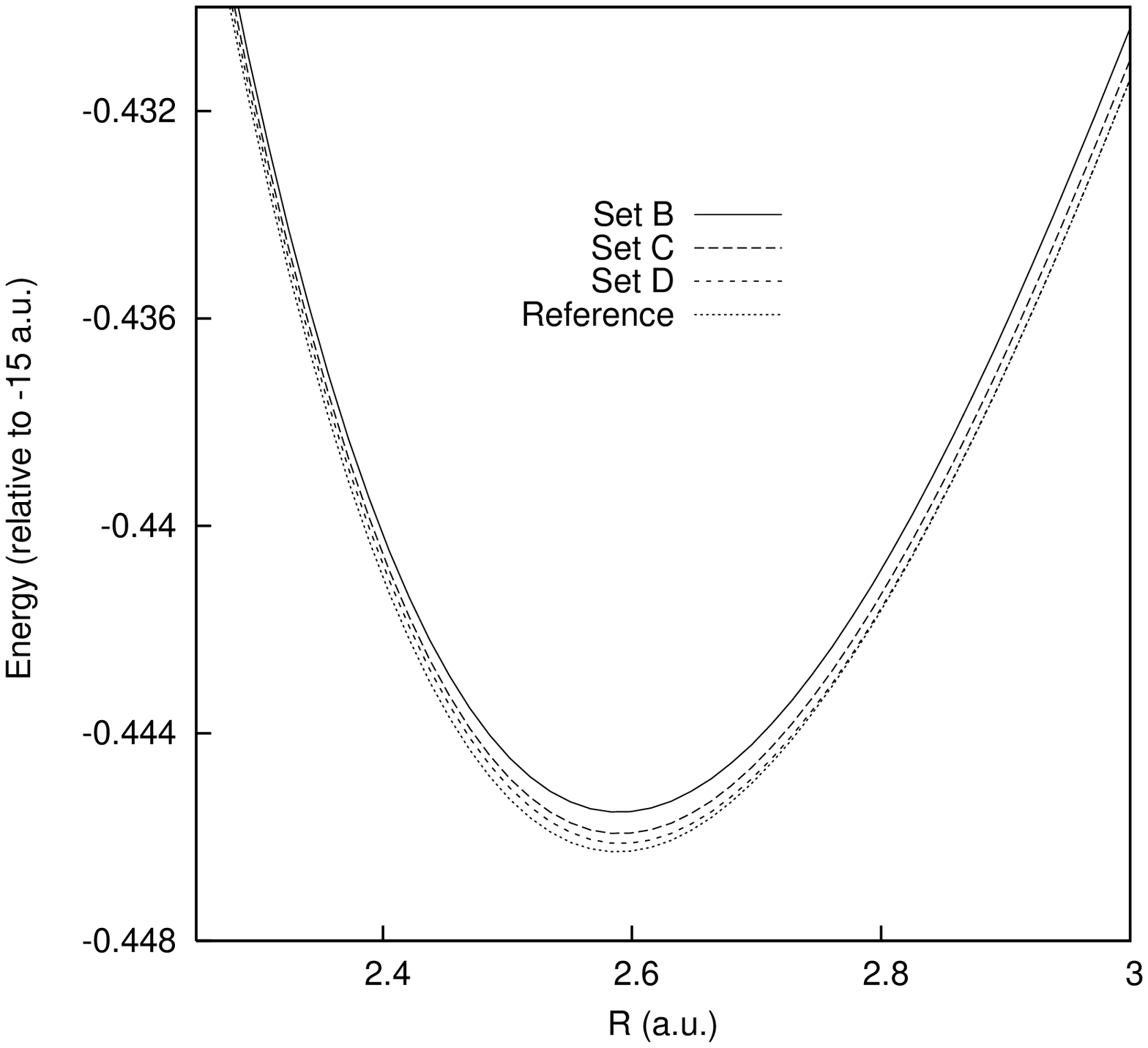}
\end{minipage}%
\caption{Cl$_2$(left panel) and HCl(right panel) potential energy curves for different grid sets.}
\end{figure}

Now a thorough comparison was made for computed eigenvalues for Cl$_2$ (at $R=4.20$ a.u.) and HCl (at
$R=$2.40 a.u.) for several grid sets \cite{roy08}. These either completely match or show an absolute maximum 
deviation of only 0.0001 a.u., for all sets, except E, where the same becomes 0.0007 a.u. Figure
I shows the potential energy curve for Cl$_2$ and HCl, for 4 Sets. Total energies for both these molecules 
are given in \cite{roy08}, for $R=3.50-5.00$ (Cl$_2$) and $R=1.60-3.10$ (HCl), at intervals of 0.1 a.u. 
\cite{roy08a}.
Clearly they faithfully reproduce reference energies for the entire range of $R$. For Cl$_2$, Set D energy 
values are quite well (higher by only 0.0001 a.u.) up to $R=4.00$ a.u., and after that shows a gradual 
tendency to deviate. Even then the maximum discrepancy is quite small (0.0007 a.u. for $R=5.00$). Sets G and 
I ($N_r=128, h_r=0.3$) either completely match with reference values or show deviations of only 0.0001 a.u. Computed energies are 
always found to be above reference values except in two occasions ($R= 4.00$ and 4.30 for Set G). Excepting 
two $R$ values, Set G 
gives exact quantitative agreement with reference results. For HCl also, very good agreement is 
observed for all three sets with reference values, with Set D performing best. 

Now that the stability and reliability of our current method is established, in Table II, we assess its 
performance for a representative set of atoms and molecules, within the LDA approximation, in terms of 
energy components
and $N$. Henceforth, in this and all other tables presented in this work, all experimental geometries 
are taken from \cite{johnson06}. Set E grid is used for all these calculations, which was found to be 
quite satisfactory for Cl$_2$, HCl. These are ordered with increasing $N$. Reference grid-DFT 
results are quoted for direct comparison. Excellent agreement is observed once again, for all of them, as 
expected. For LDA results on a more extended set of atoms and molecules, see \cite{roy08}.

\begingroup
\squeezetable
\begin{table}
\caption {\label{tab:table2}Kinetic energy, $\langle T \rangle$, potential energy, $\langle V \rangle$, 
total energy, $E$, and $N$ for selected atoms and molecules (in a.u.) within LDA. PW=Present Work.} 
\begin{ruledtabular}
\begin{tabular}{lrrrrrrrr}
System     & \multicolumn{2}{c}{$\langle T \rangle$} & \multicolumn{2}{c}{$-\langle V \rangle$} & 
\multicolumn{2}{c}{$-\langle E \rangle$}   & \multicolumn{2}{c}{$N$}  \\
\cline{2-3}  \cline{4-5} \cline{6-7} \cline{8-9} 
        & PW   & Ref.~\cite{schmidt93} &  PW & Ref.~\cite{schmidt93} &  PW & Ref.~\cite{schmidt93}  
        & PW   & Ref.~\cite{schmidt93} \\ 
\hline Na$_2$  & 0.14507   & 0.14499 & 0.52800 & 0.52791 & 0.38292  & 0.38292  
               & 1.99999   & 2.00000            \\ 
P              & 2.35430 & 2.35334 & 8.73501 & 8.73404 & 6.38070 & 6.38071 
               & 5.00000 & 4.99999                    \\     
Br             & 4.22038 & 4.22011 & 17.28157 & 17.28131 & 13.06119 & 13.06120 
               & 7.00000 & 6.99967                \\  
PH$_3$         & 4.08953  & 4.08953 & 12.27387 & 12.27383 & 8.18434   & 8.18430  
               & 8.00000 & 7.99965         \\  
SiH$_2$Cl$_2$  & 13.95036 & 13.94989 & 48.78729 & 48.78685 & 34.83693 & 34.83696 
               & 19.99999 & 20.00015 \\     
\end{tabular}                                                                               
\end{ruledtabular}
\end{table}
\endgroup
So far we have discussed only LDA results. 
It is well-known that local density functionals suffer from a number of serious problems. Hence it is 
essential to use more accurate functionals for practical calculations. Note that $E_{xc}$
accounts for two separate contributions in terms of two differences, \emph{viz.,} (i) between 
classical and quantum mechanical electron repulsion (ii) between kinetic energies of fictitious
non-interacting system and actual systems concerned. In practice, most functionals do not make
explicit attempts to incorporate this second portion. Since exact functionals are not available as yet, 
a host of fairly accurate, sophisticated, elaborate approximate functionals have been devised  
with varying degrees of complexity over the years. These have been found to be quite valuable for 
a broad range of chemical problems (see, for example, \cite{becke88a,lee88,perdew92,becke97,filatov97,
hamprecht98,proynov00,adamo02,boese02,toulouse02,tao03}, for some representative candidates). 
Commonly used functionals typically use gradient and/or Laplacian of density--the so-called generalized 
gradient approximation (GGA); there is also a class of orbital-dependent functionals, such as optimized 
effective potential approach; yet another category
combines orbital-dependent HF and an explicit density functional falling in to the variety of ``hybrid 
functionals". At this stage, we wish to extend the scope of applicability of our method by incorporating at first,
gradient corrected XC functionals (popular non-local Becke exchange \cite{becke88a} and Lee-Yang-Parr (LYP) 
correlation \cite{lee88}) are used, for illustration. For practical implementional purposes, however, an 
alternative equivalent form 
\cite{miehlich89}, containing only first derivative has been used for LYP correlation functional.
This demonstrates feasibility and viability of our current scheme in the context of non-local XC 
functionals which would be necessary for future chemical applications. Following \cite{pople92}, 
gradient-dependent functionals can be treated without evaluating density Hessians by using a 
finite-orbital basis expansion. To this end, XC contributions of KS matrix is written in 
following working form, 
\begin{equation}
F_{\mu \nu}^{XC \alpha}= \int \left[ \frac{\partial f}{\partial \rho_{\alpha}} \chi_{\mu} \chi_{\nu} +
 \left( 2 \frac{\partial f}{\partial \gamma_{\alpha \alpha}} \nabla \rho_{\alpha} + 
    \frac{\partial f} {\partial \gamma_{\alpha \beta}} \nabla \rho_{\beta} \right) 
    \cdot \nabla (\chi_{\mu} \chi_{\nu}) \right] d\rvec
\end{equation}
where $\gamma_{\alpha \alpha} = |\nabla \rho_{\alpha}|^2$, 
$\gamma_{\alpha \beta} = \nabla \rho_{\alpha} \cdot \nabla \rho_{\beta}$, $\gamma_{\beta \beta} = 
|\nabla \rho_{\beta}|^2$ and $f$ is a function only of local quantities $\rho_{\alpha}$, 
$\rho_{\beta}$ and their gradients. Non-local functionals are implemented using the Density Functional 
Repository program \cite{repository}. 

\begingroup
\squeezetable
\begin{table}      
\caption{\label{tab:table3} Variation of BLYP energy components as well as $N$ with respect to grid 
parameters for Cl$_2$ and HCl. Quantities are in a.u.}
\begin{ruledtabular}
\begin{tabular} {lrrrrrr}
    & \multicolumn{3}{c}{Cl$_2$ ($R=4.2$ a.u.)}  &  \multicolumn{3}{c}{HCl ($R=2.4$ a.u.)}      \\
\cline{2-4} \cline{5-7} 
Set                        &     A      &    B       & Ref.~\cite{schmidt93}     &   A  &   B   &  Ref.~\cite{schmidt93} \\
$N_r$                      &    64      &   128      &             &   64        &    128      &             \\
$h_r$                      &    0.3     &   0.2      &             &   0.3       &    0.2      &             \\
$\langle T \rangle$        & 11.21504   & 11.21577   &  11.21570   & 6.25431     & 6.25464     & 6.25458     \\
$\langle V^{ne}_t \rangle$ & $-$83.72582& $-$83.72695& $-$83.72685 & $-$37.29933 & $-$37.29987 & $-$37.29979 \\
$\langle V^{ee}_t \rangle$ & 31.07572   & 31.07594   &  31.07594   & 12.63884    & 12.63903    & 12.63901    \\
$\langle V \rangle$        & $-$40.98344& $-$40.98434& $-$40.98424 & $-$21.74382 & $-$21.74417 & $-$21.74411 \\
$\langle E_{el} \rangle$   & $-$41.43506& $-$41.43524& $-$41.43522 & $-$18.40618 & $-$18.40620 & $-$18.40620 \\
$\langle E \rangle$        & $-$29.76840& $-$29.76857& $-$29.76855\footnotemark[1] & 
                             $-$15.48951 & $-$15.48953 & $-$15.48953\footnotemark[2] \\
 $N$                       & 14.00006   & 14.00000   & 13.99998    & 8.00002     & 8.00000     &  8.00000    \\
\end{tabular}
\end{ruledtabular}
\footnotetext[1]{The \emph{grid-free} DFT value is $-$29.74755 a.u. \cite{schmidt93}.} 
\footnotetext[2]{The \emph{grid-free} DFT value is $-$15.48083 a.u. \cite{schmidt93}.} 
\end{table}
\endgroup
A comparison of our calculated BLYP ground-state energy 
components and $N$ with respect to variations in grid parameters are given in Table III, for Cl$_2$ and 
HCl at internuclear distances 4.2 and 2.4 a.u respectively. To make a test of our convergence,
a series of calculations were performed which produced very similar conclusions as reached in for 
LDA XC-case. From these numerical experiments, results for two selected sets are presented here, which is
sufficient to illustrate the relevant points. Reference theoretical results of \cite{schmidt93} have
been quoted once again, for comparison, employing same basis set, effective core potential within 
grid method). Some extra calculations in different grids were also done for a decent number of
molecules to monitor performance of calculated energies as well as other quantities with respect to radial and 
angular grid, \emph{viz.,} (i) $N_r, N_{\theta}, N_{\phi}$ = 96, 36, 72 (ii) $N_r, N_{\theta}, N_{\phi}$ 
= 128, 36, 72, in addition to the default grid ($N_r, N_{\theta}, N_{\phi}$ = 96, 12, 24). Three integers
here denote number of integration points in $r, \theta, \phi$ directions respectively. Overall, very similar 
results were obtained from these sets. For example, out of a total of 17 atoms and molecules, total energies 
remained unchanged up to 5th decimal place for 8 species, while in remaining cases, very slight deviations were
observed among them; the largest in total energy being 0.00064 a.u. for Na$_2$Cl$_2$ (for all others
this remains below 0.00007). However, passing from default grid to (ii) gradually improves $N$. 
Quantities considered are those same as those in Table I, \emph{viz.,} kinetic ($T$), total 
nucleus-electron attraction ($V_t^{ne}$),  total two-electron potential ($V_t^{ee}$), total potential 
($V$), electronic ($E_{el}$), total energy ($E$) and $N$. Evidently, once again, calculated results 
show excellent agreement with literature values, as found earlier in LDA case. For obvious reasons, Set 
B results are expected to be better than A; this does happen, but only marginally. Cl$_2$ shows this effect  
slightly more pronounced than that for HCl. Set B total energies show absolute deviations of 0.00002 and 
0.00000 a.u. respectively for Cl$_2$ and HCl. For all practical purposes, Set A is adequate enough for both 
of them. Note that reference \emph{grid-free} and \emph{grid}-DFT energies differ substantially from 
each other.

\begingroup
\squeezetable
\begin{table}
\caption{\label{tab:table4} Comparison of BLYP negative eigenvalues (in a.u.) of Cl$_2$, HCl with 
reference values.} 
\begin{ruledtabular}
\begin{tabular} {lccclccc}
    MO   & \multicolumn{3}{c}{Cl$_2$ ($R=4.2$ a.u.)} & MO  & \multicolumn{3}{c}{HCl ($R=2.4$ a.u.)} \\
\cline{2-4}  \cline{6-8}
Set          & A      &   B     & Ref. \cite{schmidt93} &  &  A       & B       & Ref. \cite{schmidt93} \\ 
 $2\sigma_g$ & 0.8143 & 0.8143  & 0.8143   & $2\sigma$     &  0.7707  & 0.7707  & 0.7707   \\
 $2\sigma_u$ & 0.7094 & 0.7094  & 0.7094   & $3\sigma$     &  0.4168  & 0.4167  & 0.4167   \\
 $3\sigma_g$ & 0.4170 & 0.4171  & 0.4171   & $1\pi_x$      &  0.2786  & 0.2786  & 0.2786   \\
 $1\pi_{xu}$ & 0.3405 & 0.3405  & 0.3405   & $1\pi_y$      &  0.2786  & 0.2786  & 0.2786   \\
 $1\pi_{yu}$ & 0.3405 & 0.3405  & 0.3405   &  &            &          &                    \\ 
 $1\pi_{xg}$ & 0.2778 & 0.2778  & 0.2778   &  &            &          &                    \\
 $1\pi_{yg}$ & 0.2778 & 0.2778  & 0.2778   &  &            &          &                     \\
\end{tabular}
\end{ruledtabular}
\end{table}
\endgroup
As in the LDA case, now our calculated orbital energies for Cl$_2$, HCl are compared with reference
eigenvalues within BLYP XC functional \cite{roy08a} in Table IV. In this occasion too, both molecules 
produce highly satisfactory agreement with literature values. Sets A and B results match 
\emph{completely} for \emph{all} orbital energies except the lone case of $3\sigma$ levels for both 
(with absolute deviation being only 0.0001 a.u.). Now, a thorough comparison of our computed
total energies of Cl$_2$ ($R=3.5-5.0$ a.u.) and HCl ($R=1.5-3.0$ a.u.) for BLYP XC functional was made, 
as in LDA case, for a broad range of internuclear distance, with intervals of 0.1 a.u. \cite{roy08}. It is quite 
gratifying that, for both molecules, Sets A and B results practically coincide on reference values 
for the \emph{entire} range of $R$. Cl$_2$ gives a maximum absolute deviation of 0.0001 a.u. with 
Set B and 0.0002 a.u. (only in 2 instances) with Set A. However, for HCl, the two corresponding 
deviations remain well below 0.0001 a.u. For further details, see \cite{roy08a}. 

Table V reports various energy components as well as $N$ for selected atoms, molecules 
calculated using BLYP XC functional (ordered in terms of increasing $N$ as we descend the table). 
Calculated component energies show very similar agreements with reference values as noted 
in previous tables 
and are thus omitted to avoid repetition. Atomic calculations were performed using $N_r=64, h_r=0.4$, 
whereas for molecules $N_r=128, h_r=0.3$, was used. Overall, present results agree with reference 
values excellently. Out of 5 atoms and 10 molecules, in 5 occasions, total energies are identical
as those obtained from reference theoretical method; maximum absolute deviation in total energy remains well 
below 0.0013\%. Results for more atoms and molecules are given in \cite{roy08a}.

\begingroup
\squeezetable
\begin{table}
\caption {\label{tab:table5}Comparison of BLYP energy components (in a.u.) and $N$ for selected atoms, 
molecules with reference grid-DFT results. PW=Present Work.} 
\begin{ruledtabular}
\begin{tabular}{lrrrrrrrr}
System     & \multicolumn{2}{c}{$\langle T \rangle$} & \multicolumn{2}{c}{$-\langle V \rangle$} & 
\multicolumn{2}{c}{$-\langle E \rangle$}   & \multicolumn{2}{c}{$N$}  \\
\cline{2-3}  \cline{4-5} \cline{6-7} \cline{8-9} 
        & PW   & Ref.~\cite{schmidt93} &  PW & Ref.~\cite{schmidt93} &  PW & Ref.~\cite{schmidt93}  
        & PW   & Ref.~\cite{schmidt93} \\ 
\hline 
Mg            & 0.24935  & 0.24935  &  1.06017 & 1.06017  & 0.81082  & 0.81083  & 1.99999  & 1.99999  \\
Br            & 4.27022  & 4.27043  & 17.36122 & 17.36148 & 13.09100 & 13.09105 & 7.00000  & 6.99999  \\  
MgCl$_2$      & 11.75947 & 11.75999 & 42.54049 & 42.54103 & 30.78102 & 30.78104 & 16.00004 & 15.99999 \\ 
SiH$_2$Cl$_2$ & 14.14948 & 14.14945 & 49.04463 & 49.04461 & 34.89515 & 34.89516 & 19.99999 & 20.00000 \\
\end{tabular}                                                                               
\end{ruledtabular}
\end{table}
\endgroup
Finally Table VI gives a comparison of calculated $-\epsilon_{\mathrm{HOMO}}$ (in a.u.) and atomization 
energies (in kcals/mole) with experimental results \cite{afeefy05} for select 7 molecules at their
experimental geometries \cite{johnson06}, for both LDA and BLYP XC functionals. Since reference 
theoretical results from \cite{schmidt93} are practically identical to ours (as anticipated from previous 
discussion), they are omitted here. An asterisk in experimental atomization energies denote 
298$^{\circ}$K values; otherwise they imply 0$^{\circ}$K values. Ionization energies are also reported
for another exchange functional, for the following reason. It may be remembered that simple LDA
or GGA XC potentials suffer from improper asymptotic 
\emph{long-range} behavior. Stated otherwise, whereas ground-state total energies of atoms, molecules, 
solids are estimated quite satisfactorily through these functionals, unfortunately ionization energies 
obtained from these functionals are typically off by 30-50\% of experimental values. Furthermore, 
these functionals provide a rather poor description of higher-lying excited states. At this stage, 
we note that a 
fundamental, primary objective of our proposed methodology is to extend its scope and applicability  
towards dynamical studies of atoms and molecules (particularly under the influence of strong high-intensity 
laser field through multi-photon ionization, high-order harmonic generation, photoionization,
photoemission, photodissociation, etc.) via TDDFT, that can potentially exploit the remarkable advances
already made in basis-set based DFT approaches in the past decades. This has been a highly fertile, 
fascinating area of research for more than a decade, for they (i) possess a wealth of information about
many important, fundamental physical and chemical phenomena (often counter-intuitive) occurring in such 
systems (ii) could lead to diverse practical applications in
various branches of science, engineering (iii) pose enormous complications and challenges for both 
experiment and theory (some representative references are, for example, \cite{brabec00,udem02,
stapelfeldt03,baltuska03,posthumus04,gohle05}). Of late, activity in this field has dramatically increased, 
as evident from an enormous number of sophisticated experimental and theoretical works reported in the 
literature. Theoretically, while there exists a few approaches to deal with such systems (mostly within 
single active electron or frozen-core approximation), dependable \emph{ab initio} calculations have been 
very much limited for atoms and even more for molecules, so much so that accurate calculations of even H$_2$
molecule is a significantly challenging task, not to speak of polyatomic molecules, in general. Moreover 
whereas remarkable advancements has been made for atomic case, molecular situation, nowadays, is in its 
embryonic stage only. Our long term goal is to formulate a general non-perturbative TDDFT-based formalism for 
dynamical studies of such species using an LCAO-MO prescription for solution of the TDKS 
equation, rather than a grid-discretization of molecular KS equation, as has been traditionally employed. 

A necessary prerequisite for such dynamical  
studies is that both ionization energies as well as higher levels be approximated more accurately. 
The modified Leeuwen-Baerends (LB) potential \cite{leeuwen94, schipper01}, $v_{xc\sigma}^{LB\alpha} 
(\alpha, \beta: \rvec)$, which contains two empirical parameters, seems to be a very good choice in this
game (see, for example, \cite{chu05}, and the references therein). This is conveniently written as, 
\begin{equation}
v_{xc\sigma}^{LB\alpha} (\alpha, \beta: \rvec) = \alpha v_{x\sigma}^{LDA} (\rvec) + v_{c\sigma}^{LDA} 
(\rvec) +
\frac{\beta x_{\sigma}^2 (\rvec) \rho_{\sigma}^{1/3} (\rvec)} 
{1+3\beta x_{\sigma}(\rvec) \mathrm{ln} \{x_{\sigma}(\rvec)+[x_{\sigma}^2 (\rvec)+1]^{1/2}\}}
\end{equation}
Here $\sigma$ signifies up,down spins and last term containing gradient correction bears some resemblance 
with the exchange energy density functional of \cite{becke88a}, $x_{\sigma}(\rvec) = |\nabla 
\rho_{\sigma}(\rvec)|[\rho_{\sigma}(\rvec)]^{-4/3}$ is simply a dimensionless quantity, $\alpha=1.19, 
\beta=0.01$. The desired long-range property is satisfied properly, i.e., $v_{xc\sigma}^{LB\alpha} 
(\rvec) \rightarrow -1/r, r \rightarrow \infty.$ Ionization energies obtained from LBVWN (LB exchange, 
VWN correlation) combination of XC functionals are presented in column 4 to demonstrate its noticeable
superiority
over LDA or BLYP counterpart. It is clear that, for all these species considered, LDA ionization 
energies are consistently lower than corresponding BLYP values, whereas LBVWN values are 
substantially lower and presumably more accurate (this feature will be very important for TDDFT 
calculations in future which requires accurate higher-lying states) than both these two functionals. 
Now, computed LDA, BLYP, atomization energies 
are also compared with experimental results (including zero-point vibrational correction and relativistic
effects) in columns 6,7, which shows considerable deviation from experimental results. Surprisingly, LDA results 
are apparently better than their BLYP counterparts, in several occasions. However this observation 
should not be taken too optimistically to favor the former over the latter, for there could be some 
cancellation of errors in LDA case. Moreover, several other factors, such as use of more appropriate 
pseudopotential, basis set, etc., should be taken into consideration. Large deviations are observed in 
atomization energies, which could be also due to the above factors. However, such discrepancies are not 
so uncommon in DFT, even in more elaborate and extended basis sets within all-electron calculation 
as well (see, for example, \cite{cafiero06}). In any case, this is an evolving process and does not 
directly interfere with the main objective of this work.

\begingroup
\squeezetable
\begin{table}
\caption {\label{tab:table6} Negative HOMO energies, $-\epsilon_{\mathrm{HOMO}}$ (in a.u.) and 
atomization energies (kcals/mol) for a cross-section of molecules. LDA, LBVWN (LB+VWN), BLYP results are
compared with experiment \cite{afeefy05}. An asterisk indicates 298$^{\circ}$K values. Otherwise, 
0$^{\circ}$K results are given.} 
\begin{ruledtabular}
\begin{tabular}{lccccccc}
Molecule & \multicolumn{4}{c}{$-\epsilon_{\mathrm{HOMO}}$ (a.u.)} & \multicolumn{3}{c}
{Atomization energy (kcals/mol)}  \\
\cline{2-5}  \cline{6-8}  
              &  LDA    & BLYP    &  LBVWN  & Expt.~\cite{afeefy05}    &  LDA     &  BLYP    &    
Expt.~\cite{afeefy05}   \\
\hline 
Al$_2$        & 0.1407  & 0.1400  & 0.2371  & 0.1984  & 22.92    &  21.42   &    37.0    \\
HI            & 0.2518  & 0.2432  & 0.3824  & 0.3817  & 82.82    &  72.07   &    45.8*   \\
SiH$_4$       & 0.3188  & 0.3156  & 0.4624  & 0.4042  & 339.43   &  312.02  &    302.6   \\
S$_2$         & 0.2007  & 0.2023  & 0.3443  & 0.3438  & 56.75    &  52.47   &    100.8   \\
SiCl$_2$      & 0.2514  & 0.2448  & 0.3909  & 0.3804  & 190.40   & 155.11   &    202.7   \\
P$_4$         & 0.2712  & 0.2575  & 0.3964  & 0.3432  & 200.77   & 142.99   &    285.9   \\
PCl$_5$       & 0.2825  & 0.2722  & 0.4397  & 0.3748  & 246.22   & 145.33   &    303.2   \\
\end{tabular}                                                                              
\end{ruledtabular}
\end{table}
\endgroup

\section{Future and Outlook}
We have demonstrated the plausibility and feasibility of simple CCGs in atomic/molecular domain within 
the context of Gaussian-based LCAO approach to DFT. For a modest number of atoms, molecules, this has
been found to be extremely accurate and dependable; producing practically identical results with those 
obtained from other grid-based/grid-free quantum chemistry programs available in the literature.
Atom-centered localized basis set, MOs, two-body potentials have been constructed on a 3D cubic 
box directly. Illustrative results have been presented for local and non-local XC density functionals.
Classical Hartree potential was obtained through a Fourier convolution technique, accurately and 
efficiently in the real grid. No auxiliary basis set was invoked and Hay-Wadt-type pseudopotentials was
used to describe core electrons. The method has been overwhelmingly successful in predicting static
properties, such as total energy and component energies, HOMO energies, potential energy curves, 
orbital energies, atomization energies. The results were found to be variationally bounded. 

This works provides the necessary springboard to explore the possibility of investigating real-time
dynamical studies of many-electron systems which constitutes one of our next immediate future objective.
In particular, our interest lies in the laser-atom/molecule interactions in intense/superintense regime,
which are inherently non-perturbative, non-linear phenomena. Although a number of theoretical methods 
have been developed for their studies, TDDFT seems to be the most promising among them. Within TDDFT, 
however, only real-space methods have been employed in a number of such problems with some success. 
However, to our knowledge, no attempt has been made so 
far to exploit a Gaussian-based DFT approach (which has been so profoundly successful for a multitude of 
static problems over the past several decades) in real-time dynamical regime. The overwhelming success of 
this work, as illustrated here, encourages to venture into such an undertaking and it is hoped that this 
will be equally fruitful in our future dynamical studies. Incorporation of LB exchange potential was a 
necessary first step in developing a basic workable solution for this. Introduction of more appropriate 
pseudopotentials devoted to density functional methods as well as more elaborate, extended, sophisticated 
basis sets would be highly desirable and worthwhile. Application of this method to other interesting chemical 
problems such as weakly bonded systems including H-bonded molecules, clusters and larger systems would 
further consolidate its success. A systematic analogous investigation of its performance on all-electron 
systems would also be interesting.

\section{Acknowledgment}
I express my sincere gratitude to the Editor, Prof. F.~Columbus, for giving me this opportunity to present my 
work in this edition. I sincerely thank Nova Publishers, NY, USA, for their full, generous support, 
and extending the deadline for submitting manuscript. I am indebted to professors,
D.~Neuhauser, E.~Proynov, S.~I.~Chu and Z.~Zhou, for numerous 
discussions, comments, suggestions. Warm hospitality provided by the Univ. of California, Los 
Angeles and Univ. of Kansas, Lawrence, USA is gratefully acknowledged. This work could not have 
been possible without the support of my colleagues at IISER, Kolkata, India, for providing a cordial,
friendly atmosphere, and I thank all of them very much.

\bibliography{refn.bib}
\bibliographystyle{unsrt}
\end{document}